# 5

# Parallel MATLAB Techniques

Ashok Krishnamurthy, Siddharth Samsi and Vijay Gadepally
*Ohio Supercomputer Center and Ohio State University*
*U.S.A.*

## 1. Introduction

MATLAB is one of the most widely used languages in technical computing. Computational scientists and engineers in many areas use MATLAB to rapidly prototype and test computational algorithms because of the scripting language, integrated user interface and extensive support for numerical libraries and toolboxes. In the areas of signal and image processing, MATLAB can be regarded as the de facto language of choice for algorithm development. However, the limitations of desktop MATLAB are becoming an issue with the rapid growth in the complexity of the algorithms and the size of the datasets. Often, users require instant access to simulation results (compute bound users) and/or the ability to simulate large data sets (memory bound users). Many such limitations can be readily addressed using the many varieties of parallel MATLAB that are now available (Choy & Edelman, 2005; Krishnamurthy et al., 2007). In the past 5 years, a number of alternative parallel MATLAB approaches have been developed, each with its own unique set of features and limitations (Interactive Supercomputing, 2009; Mathworks, 2009; MIT Lincoln Laboratories, 2009; Ohio Supercomputer Center, 2009).

In this chapter, we show why parallel MATLAB is useful, provide a comparison of the different parallel MATLAB choices, and describe a number of applications in Signal and Image Processing: Audio Signal Processing, Synthetic Aperture Radar (SAR) Processing and Superconducting Quantum Interference Filters (SQIFs). Each of these applications have been parallelized using different methods (Task parallel and Data parallel techniques). The applications presented may be considered representative of type of problems faced by signal and image processing researchers. This chapter will also strive to serve as a guide to new signal and image processing parallel programmers, by suggesting a parallelization strategy that can be employed when developing a general parallel algorithm. The objective of this chapter is to help signal and image processing algorithm developers understand the advantages of using parallel MATLAB to tackle larger problems while staying within the powerful environment of MATLAB.

## 2. Parallel MATLAB overview

The need for parallel MATLAB is presented in (Choy & Edelman, 2005) and the need for parallelizing MATLAB in particular can be summarized as follows:

1. MATLAB is user friendly
2. MATLAB is popular

In a survey of parallel MATLAB technologies, nearly 27 parallel MATLAB technologies were discovered. Many of these technologies are defunct, while many of these technologies are actively under development, with a large user base and active developer base. In our experience, three of these technologies stand out in terms of such factors.
In this section, we introduce three alternatives for parallel computing using MATLAB. The technologies we will be looking at are: pMATLAB+bcMPI, the Parallel Computing Toolbox (PCT) with MATLAB Distributed Computing Server and Star-P.

### 2.1 bcMPI

Traditionally, researchers have used MatlabMPI (Kepner & Ahalt, 2003) for parallel computing in MATLAB. bcMPI is an open source software library that is an alternative to MatlabMPI and is geared towards large, shared supercomputer centers. The bcMPI library was developed at the Ohio Supercomputer Center (OSC) to provide an efficient, scalable communication mechanism for parallel computing in MATLAB while maintaining compatibility with the MatlabMPI API (Hudak et al., 2007). The bcMPI package consists of an interface to the MPICH or OpenMPI library and a toolbox for MATLAB that implements a subset of the MatlabMPI API calls. bcMPI has been developed primarily on the Linux platform, but it has also been tested on the Mac OS-X, NetBSD and IA32 platforms. At its core, bcMPI is a C library that supports a collection of MATLAB and Octave data types. The bcMPI software architecture is as shown below:

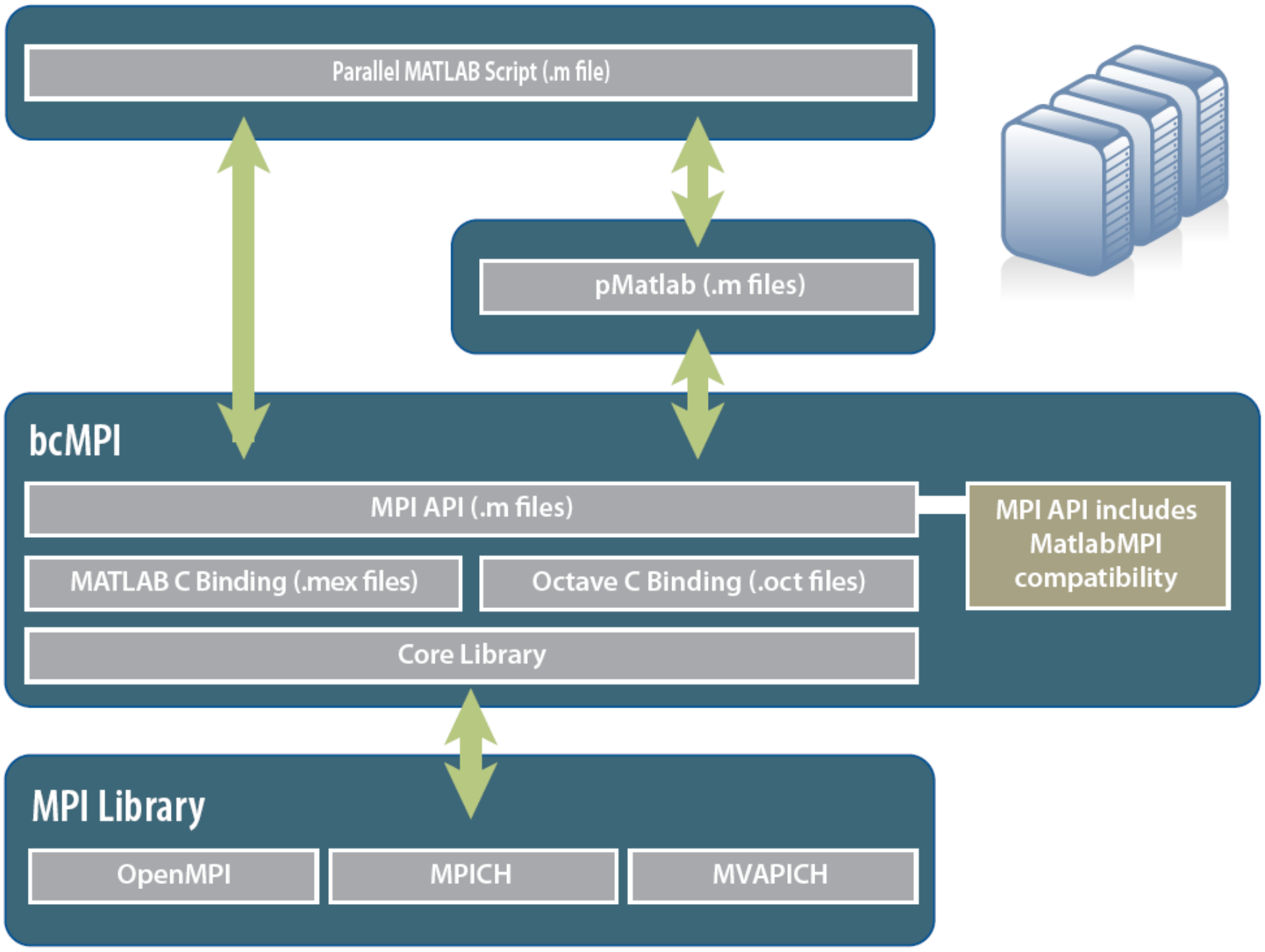

Fig. 1. bcMPI Architecture

Figure 1 illustrates the relationship between the vaious layers in the bcMPI architecture. The bcMPI library provides functions for synchronous as well as asynchronous communication between the MATLAB processes. It supports basic MPI functions as well as collective operations such as MPI_Reduce, MPI_Gather and MPI_Barrier. bcMPI also has an efficient implementation of the MPI_Broadcast function using the underlying MPI library. bcMPI has the advantage that it can use any MPI libraries even thought it has been tested actively with the OpenMPI and MPICH libraries. bcMPI interfaces with pMATLAB (a parallel MATLAB extension developed by MIT Lincoln Laboratory) (Bliss & Kepner, 2007) for distributed data processing. The combination of pMATLAB and bcMPI is denoted as pMATLAB+bcMPI. pMATLAB+bcMPI uses a layer of abstraction beyond traditional MPI calls and reduces programming complexity. With this combination, a user would not need to use explicit message passing calls to distribute data, as the pMATLAB application would perform these actions.

### 2.2 Parallel computing toolbox

The Parallel Computing Toolbox (PCT) along with the MATLAB Distributed Computing Server (MDCS) are commercial products offered by The MathWorks Inc. While the core MATLAB software itself supports multithreading, the PCT provides functionality to run MATLAB code on multicore systems and clusters. The PCT provides functions for parallel for-loop execution, creation/manipulation of distributed arrays as well as message passing functions for implementing fine grained parallel algorithms.

The MATLAB Distributed Computing Server (MDCS) gives the ability to scale parallel algorithms to larger cluster sizes. The MDCS consists of the MATLAB Worker processes that run on a cluster and is responsible for parallel code execution and process control. Figure 2 illustrates the architecture of PCT and MDCS

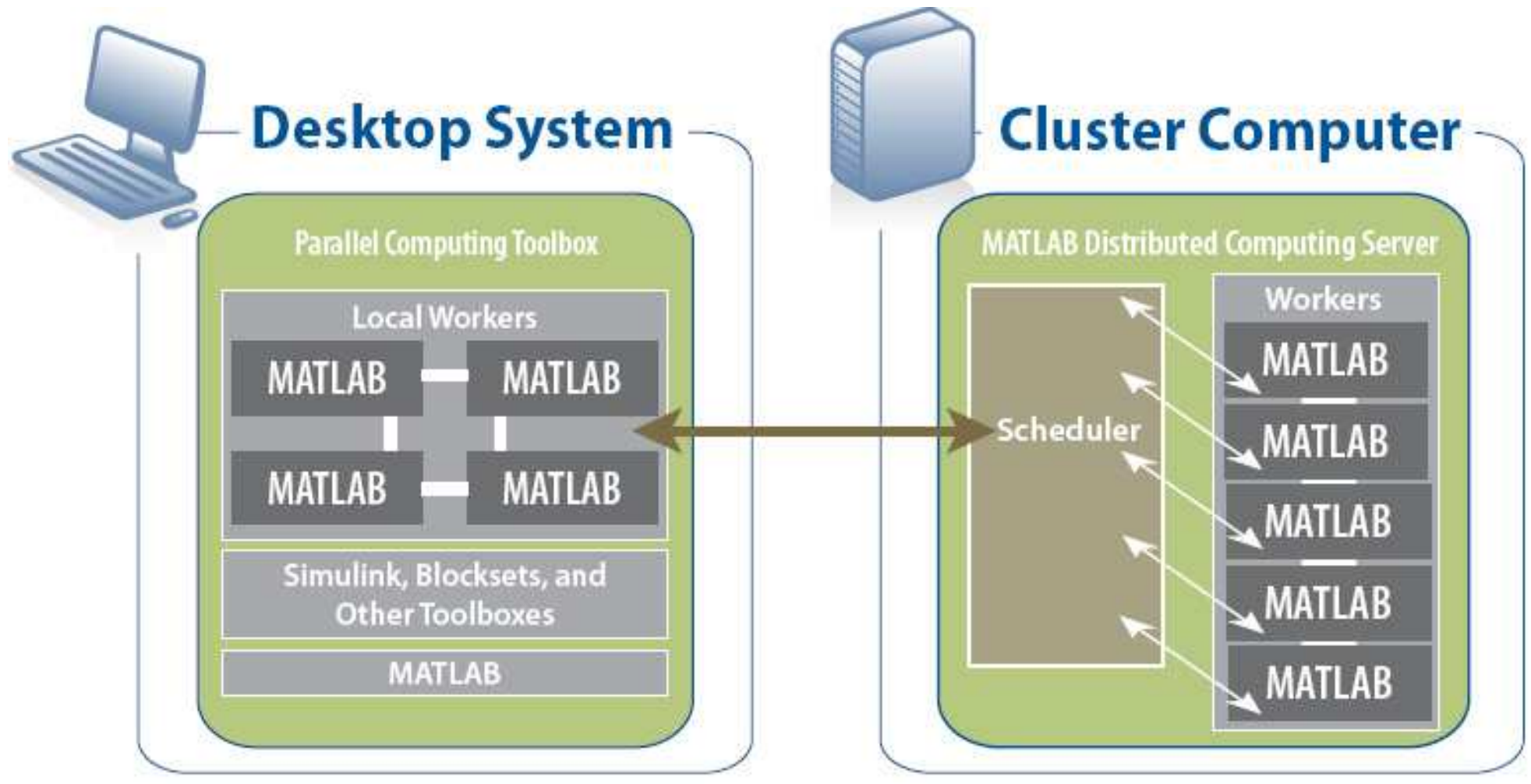

Fig. 2. The Parallel Computing Toolbox and MATLAB Distributed Computing Server

The PCT also allows users to run up to 8 MATLAB Labs or Workers on a single machine. This enables interactive development and debugging of parallel code from the desktop. After parallel code has been developed, it can be scaled up to much larger number of Worker or Labs in conjunction with the MDCS.

### 2.3 Star-P

Star-P is a client-server parallel computing platform for MATLAB available from Interactive Supercomputing. The architecture of Star-P is shown in the figure below:

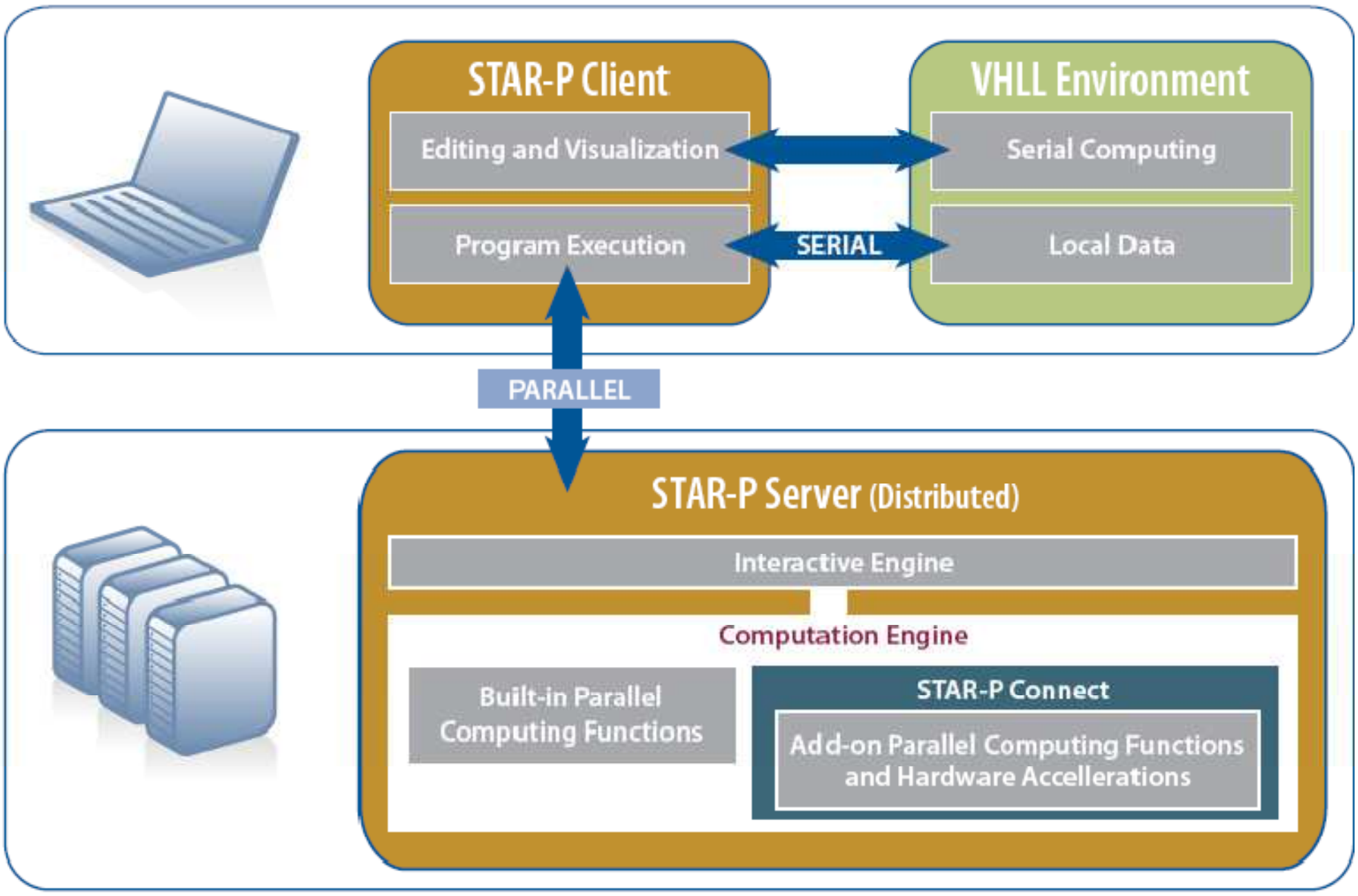

Fig. 3. Star-P Architecure

Figure 3 illustrates the structure of Star-P, and difference between the Star-P client and server. Star-P supports fine grained parallel as well as embarrassingly parallel modes of operation (these modes of operation are discussed in the next section). The biggest advantage offered by Star-P is that it eliminates the need for the developer to use explicit Message Passing Interface (MPI) message passing calls for communicating between the back-end processes. By using the "*p" construct, users can simply indicate the variables or data that are meant to be distributed over the back-end processes.

## 3. Parallel programming

The goals of most parallel computing algorithms include either reduction in computation time (for compute bound users) or analysis of larger data sets/parameters sweeps (for memory bound users) or some combination of both. This can also be described as a capability or capacity problem. In many cases, analysis involves small data sets, but the time required to analyze the desired data along with a wide enough parameter sweep on the same can make it impractical to run such analyses. In such (and a variety of other) cases, the use of parallel computing techniques can enable researchers to employ large numbers of processors to run comprehensive analyses in a reasonable amount of time. For example, the reconstruction of micro-CT images to generate 3D models may take up to 13 hours on a single machine. This can be reduced significantly simply by running the reconstruction algorithm in a parallel fashion. In other applications such as automated image analysis in medicine, the data sets tend to be large, with individual images ranging in the multiple gigabytes. In such cases, it may not be possible to load the data into memory and analyze the images. At the Ohio Supercomputer Center, we have observed that reading in images as large as 60000x60000 pixels in resolution on a single machine with 32GB RAM can take upwards of 30 minutes. Running simple algorithms on such large images becomes impractical, with software stability also becoming a concern. Furthermore, given many such images for analysis, the time required to run a single analysis on all images becomes impractical and parameter sweep studies become constrained by time. High resolution images obtained from tissue biopsies can be as large as 30-40GB each, and with the existing software and hardware limitation it is not possible to read in entire images on a single processor, thus leading to the problem of capability. In such cases, a simple solution is to use parallel computing in MATLAB to process parts of the image on separate processors to address problem.

Broadly, parallel algorithms can be divided into two categories: Task Parallel and Data Parallel. Task parallel (or Embarrassingly Parallel) algorithms take advantage of the fact that multiple processors can work on the same problem without communicating with each other. Typical cases of such algorithms include Monte Carlo simulations where the order of computations in a large loop are independent of each other and can be performed in any order without affecting the results. Similarly, another application ideal for task parallelism involves processing multiple datasets using the same algorithm. In such cases multiple processors can analyze subsets of the data simultaneously without the need for inter-processor communication. Data parallel (or Fine Grained Parallel) algorithms typically involve some inter-processor communication. In such algorithms the data to the analyzed is typically too large to be analyzed on a single processor. Parallel computing paradigms are used to distribute the data across processors and each processor works on a smaller chunk of the same data. In such cases, there may be some communication required between different processors that involve exchange of data to address boundary conditions. For example, a 2-D FFT of a large matrix can be carried out in a parallel fashion by splitting up the matrix across multiple processors. Based on how the data is distributed, each processor needs a small amount of data from its neighbour to complete the computations.

The maximum speed up (ratio of runtime before parallelization to runtime after parallelization) is discussed in (Amdahl, 1967). The maximum observable speedup is limited by the percent of the application that can be parallelized. The maximum percentage of the application that can be parallelized is determined by the percentage of code that must be run serially. This serial execution requirement is often due to data dependencies present in the code, or complications that may arise due to parallelization. It is important that a parallel programmer determine the maximum speed up before beginning parallelization. In certain applications regardless of parallelization technique, the required speedup may not be attainable.

In the next section, we discuss three applications that help illustrate the different types of parallel algorithms discussed here. Two of the applications being considered can be parallelized using either the task parallel or data parallel technique. One of the presented applications can be parallelized using both techniques, and a comparison is provided.

## 4. Application development

In this section, three applications will be presented. The aim of this section is to give real life examples of the discussed parallel MATLAB technologies in action (Krishnamurthy et al., 2008). Additionally, this section will suggest methods by which parallel MATLAB programmers can approach a given parallelization problem. The following applications have been parallelized using (1) Task Parallel (Embarrassingly Parallel) and/or (2) Data Parallel (Fine Grained Parallel) techniques.

For each of the applications developed we will concentrate on the following:

1. Application Background
   This section will give background information on the application. This section is intended to show readers the variety of problems that can be tackled using parallel MATLAB.
2. Parallelization Strategy
   This section will describe the strategy employed when parallelizing the application. Additionally, specific code examples from the serial code, and our parallel code for the same will be shown.
3. Results
   This section will demonstrate the results obtained through parallelization. This section is important in illustrating the computational benefits possible through parallelization.

### 4.1 Acoustic signal processing

#### 4.1.1 Application background:

Acoustic signal processing on a battlefield primarily involves detection and classification of ground vehicles. By using an array of active sensors, signatures of passing objects can be collected for target detection, tracking, localization and identification. One of the major components of using such sensor networks is the ability of the sensors to perform self-localization. Self-localization can be affected by environmental characteristics such as the terrain, wind speed, etc. An application developed by the U.S. Army Research Laboratory, GRAPE, consists of a Graphical User Interface (GUI) for running acoustic signal processing algorithms in parallel on a cluster.

In recent experiments, several gigabytes of data were collected in 3-minute intervals. Processing each data file takes over a minute. A number of different algorithms are used to estimate the time of arrival of the acoustic signals. If the number of analysis algorithms applied to the data is increased, the processing and analysis time increases correspondingly. In order to achieve near real-time response, the data was processed in a parallel fashion in MATLAB. Since each data file can be processed independently of others, the parallelization approach was to split up processing of individual data files across multiple processors.

#### 4.1.2 Parallelization strategy:

It was determined that this particular application could be parallellized using task parallel (Embarrassingly Parallel) techniques. Using the MATLAB profiler, it was determined that the majority of computation time was spent in the execution of a function called *process_audio()*. It was further determined that the data generated by this function was not used in other places (data independence). Thus, a task parallel approach was employed. The function *process_audio()* takes a data structure as an input. One of the fields in this structure is *nFiles*, which describes the number of files to be run by the *process_audio()* function. A distributed array was used to distribute these indices across multiple processors. The code before and after parallelization are as follows:

```
%%%%SERIAL CODE%%%%               %%%%PARALLEL CODE%%%%

                                  Alocal = A;
                                  indlocal = local(indices)
out = process_audio(A)            indlocal = indlocal + [(Nfiles-size(indlocal,2)+1):Nfiles];
                                  Alocal.fname = A.fname(indlocal)
                                  out = process_audio(Alocal)
```

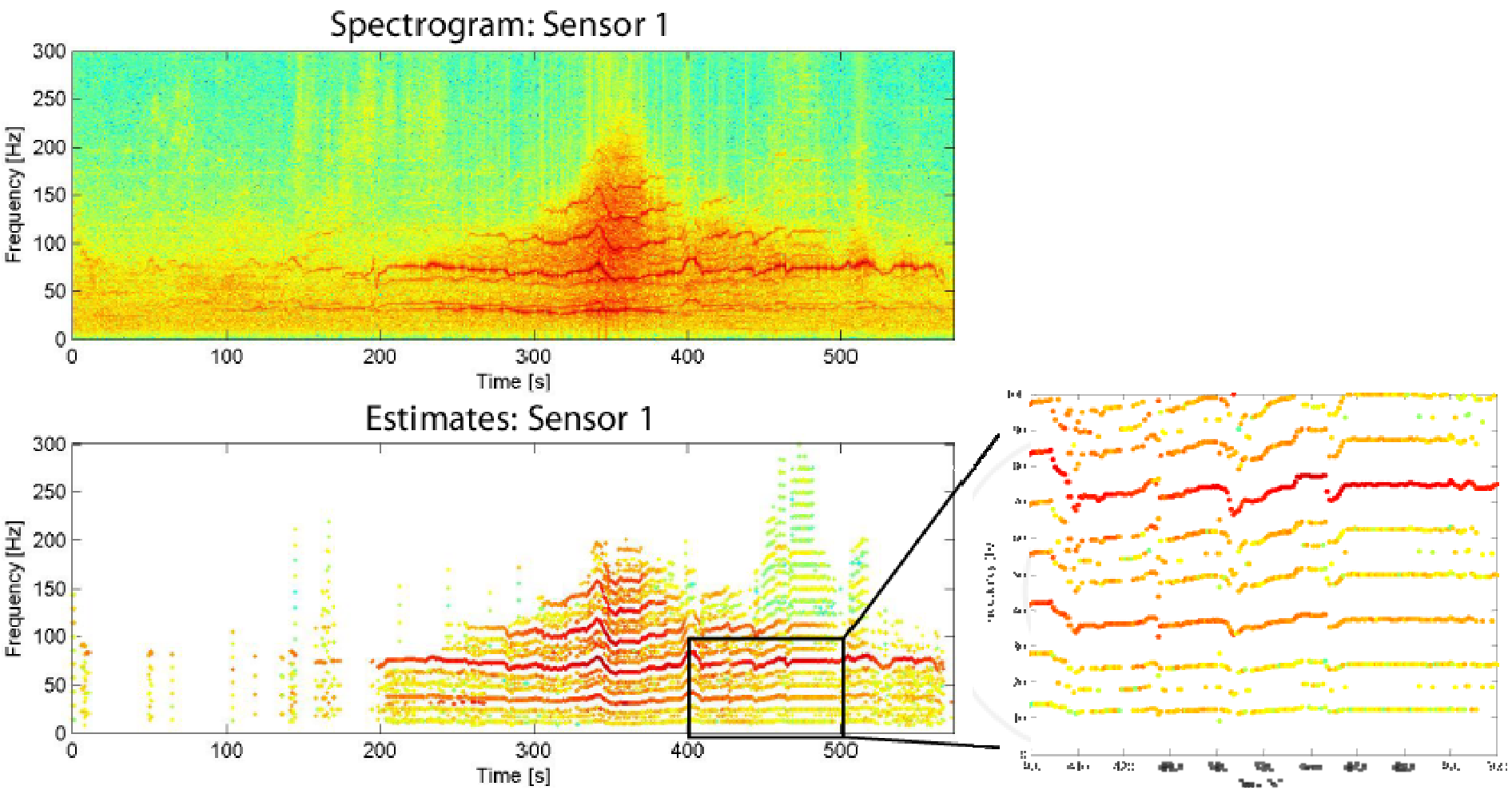

Fig. 4. Vehicle signature identified via signal processing

Fig. 5. pMATLAB parallelization of acoustic signal processing application

#### 4.1.3 Results:

Results from parallelization of the GRAPE code using MDCS, bcMPI, and Star-P are presented below. In the following graphs, the primary (left) vertical axis corresponds to the total time taken by the *process_audio()* function to complete analysis on 63 data files. The secondary (right) axis displays the speedup obtained when running on multiple processors. It is also interesting to note that the modifications required to parallelize the code represented an increase of less than 1% in Source Lines of Code (SLOC).

Results for the MDCS and bcMPI tests were obtained on the Ohio Supercomputer Center's Pentium 4 cluster using an InfiniBand interconnection network. Results for the Star-P tests were obtained on the Ohio Supercomputer Center's IBM 1350 Cluster, with nodes containing dual 2.2 GHz Opteron processors, 4 GB RAM and an InfiniBand interconnection network. As the parallelization strategy is a task parallel solution that incurs no interprocess communication, a nearly linear speed-up is observed for each of the parallel MATLAB tools. It is also clear that parallel MATLAB can aid greatly in returning a timely solution to the user.

From the above results (Figures 6, 7 and 8), it is also clear that the three technologies give nearly the same speedup for a given code set. For the remaining applications, results are shown using pMATLAB+bcMPI, and similar results can be obtained by using any of the presnted tools.

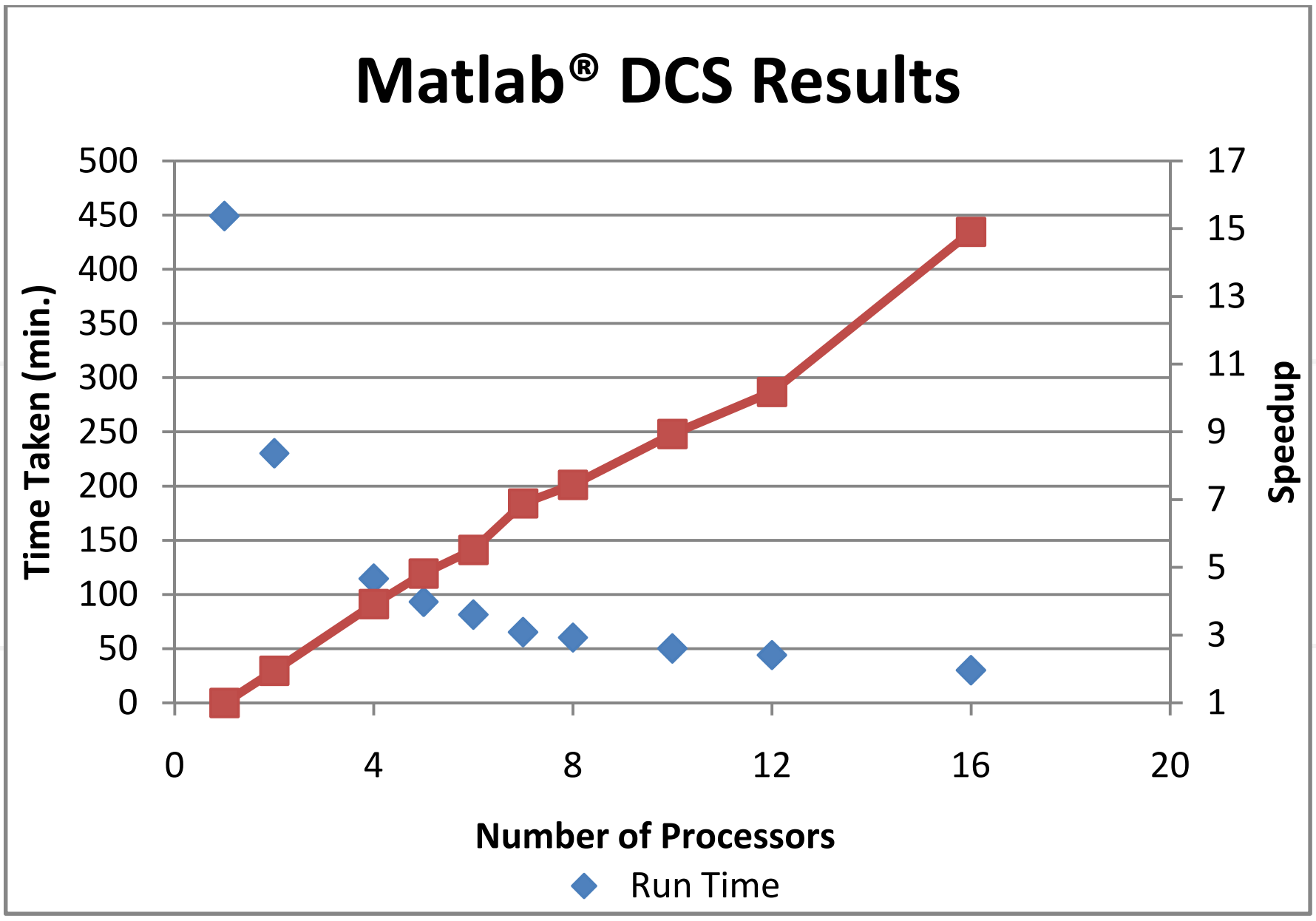

Fig. 6. MDCS Results

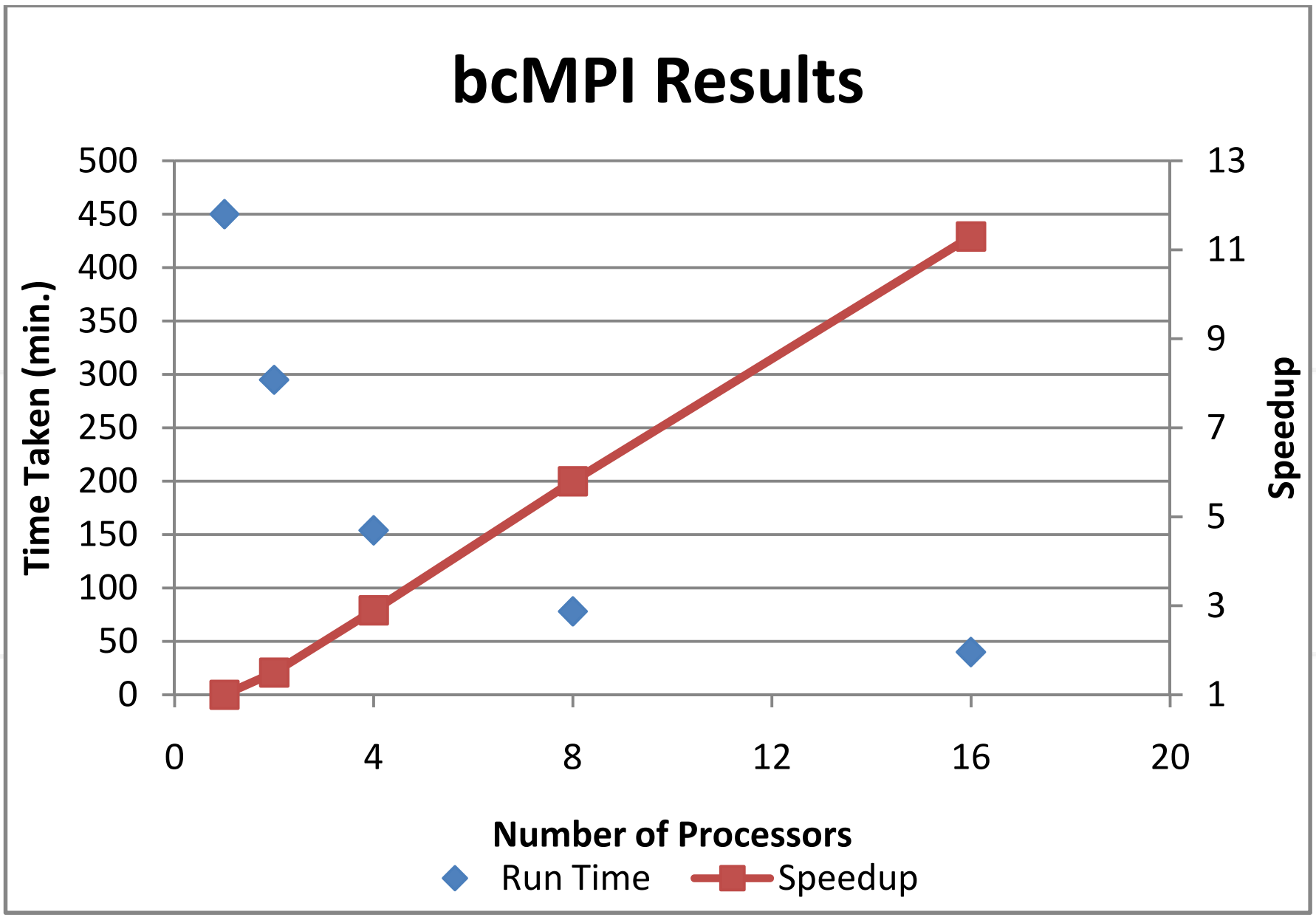

Fig. 7. bcMPI Results

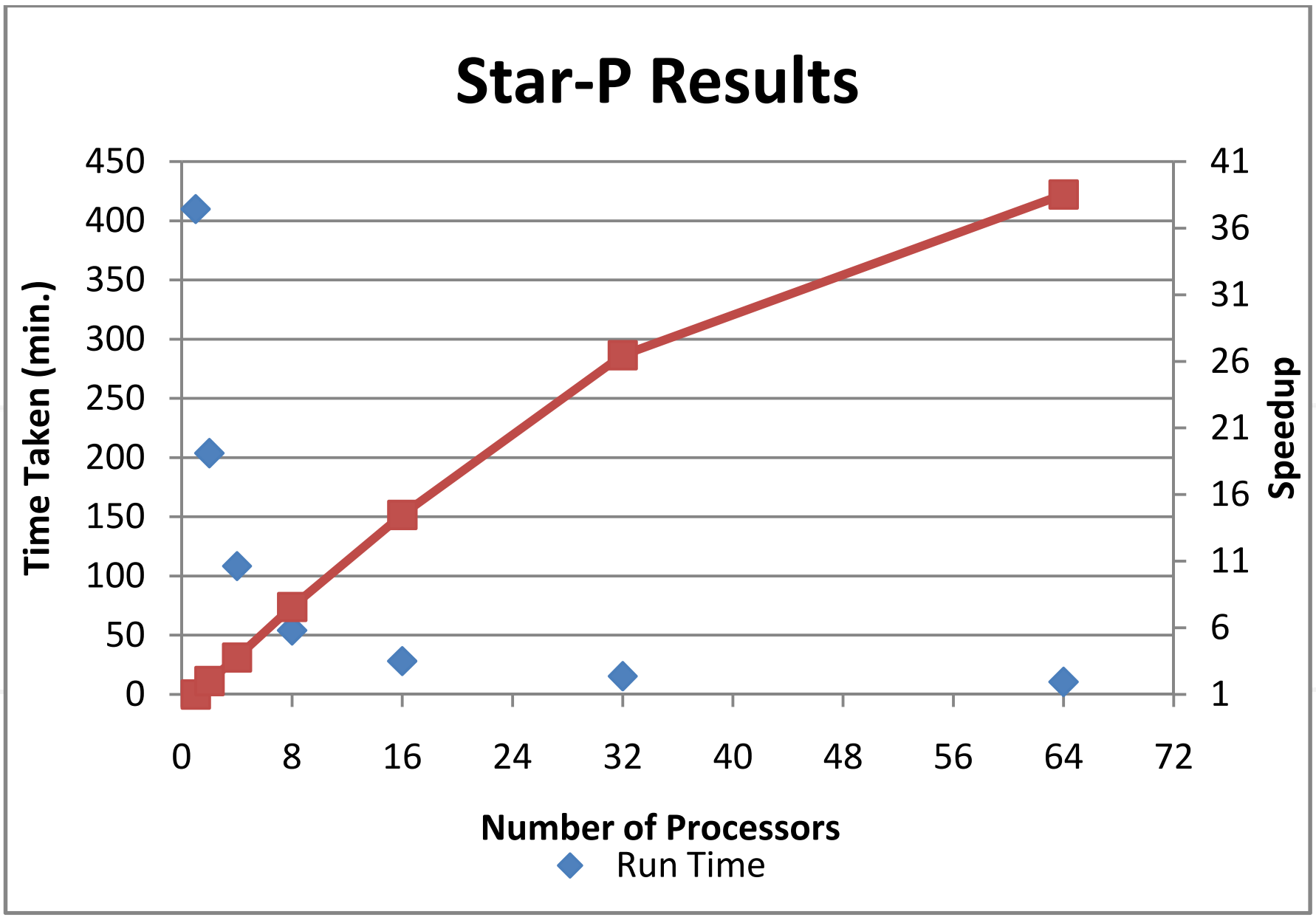

Fig. 8. Star-P Results

### 4.2 Synthetic aperture radar

#### 4.2.1 Application background:

The Third Scalable Synthetic Compact Application (SSCA #3) benchmark (Bader et al., 2006), from the DARPA HPCS Program, performs Synthetic Aperture Radar (SAR) processing. SAR processing creates a composite image of the ground from signals generated by a moving airborne radar platform. It is a computationally intense process, requiring image processing and extensive file IO. Such applications are of importance for Signal and Image Processing engineers, and the computations performed by the SSCA #3 application are representative of common techniques employed by engineers.

#### 4.2.2 Parallelization strategy:

In order to parallelize SSCA #3, the MATLAB profiler was run on the serial implementation. The profiler showed that approximately 67.5% of the time required for computation is spent in the image formation function of Kernel 1 (K1). Parallelization techniques were then applied to the function *formImage* in K1. Within *formImage*, the function *genSARimage* is responsible for the computationally intense task of creating the SAR image. *genSARimage* consists of two parts, namely, the interpolation loop and the 2D Inverse Fourier Transform. Both of these parts were parallelized through the creation of distributed matrices and then executed via pMatlab/bcMPI.

A code example is presented in Figure 9 showing the serial and parallel versions of one code segment from the function *genSARimage*. It is interesting to note that nearly 67.5% of the code was parallelized by adding approximately 5.5% to the SLOC. In the sequential code on the left of Figure 9, the matrix, *F*, needs to be divided among the processor cores to parallelize the computation. In the parallel code on the right of Figure 9, the upper shaded region shows the creation of a distributed version of the matrix, *pF*, which is distributed as contiguous blocks of columns distributed across all processors. The code within the loop remains functionally equivalent, with the parallel version altered so that each processor core processes its local part of the global array. The lower shaded region shows a pMatlab *transpose_grid* (Bliss and Kepner, 2006) operation, which performs all-to-all communication to change *pF* from a column distributed matrix to a row distributed matrix in order to compute the following inverse FFT in parallel. Finally, in the lowest shaded region, the entire pF array is aggregated back on a single processor using the pMatlab *agg* command.

```
%%%%SERIAL CODE%%%%


F = single(zeros(nx, m));








spatial=ftshift(ifft(ifft(fftshift(F),[],2))
```

```
%%%%PARALLEL CODE%%%%

kxlocal=kx(:,(myrank*pFlocalsize(2)+1):(myrank+1)*pFlocalsize(2))
KXlocal=KX(:,(myrank*pFlocalsize(2)+1):(myrank+1)*pFlocalsize(2))
fsmlocal=fsm(:,(myrank*pFlocalsize(2)+1):(myrank+1)*pFlocalsize(2)
m = length((myrank*pFlocalsize(2) +1):(myrank+1)*pFlocalsize(2))
pFmap = map([1 Ncpus], {}, [0:Ncpus-1])
pF = zeros(nx,m,pFmap);
pFlocal = ifft(pFlocal, [],2);
pF = put_local(pF, pFlocal);

Z = transpose_grid(pF);
clear pF, pFlocal;
Zlocal = local(Z);
Zlocal = ifft(Zlocal, [],1);
Z = put_local(Z,Zlocal);
Z = agg(Z);
spatial = abs(Z)';
```

Fig. 9. pMATLAB parallelization of image formation kernel

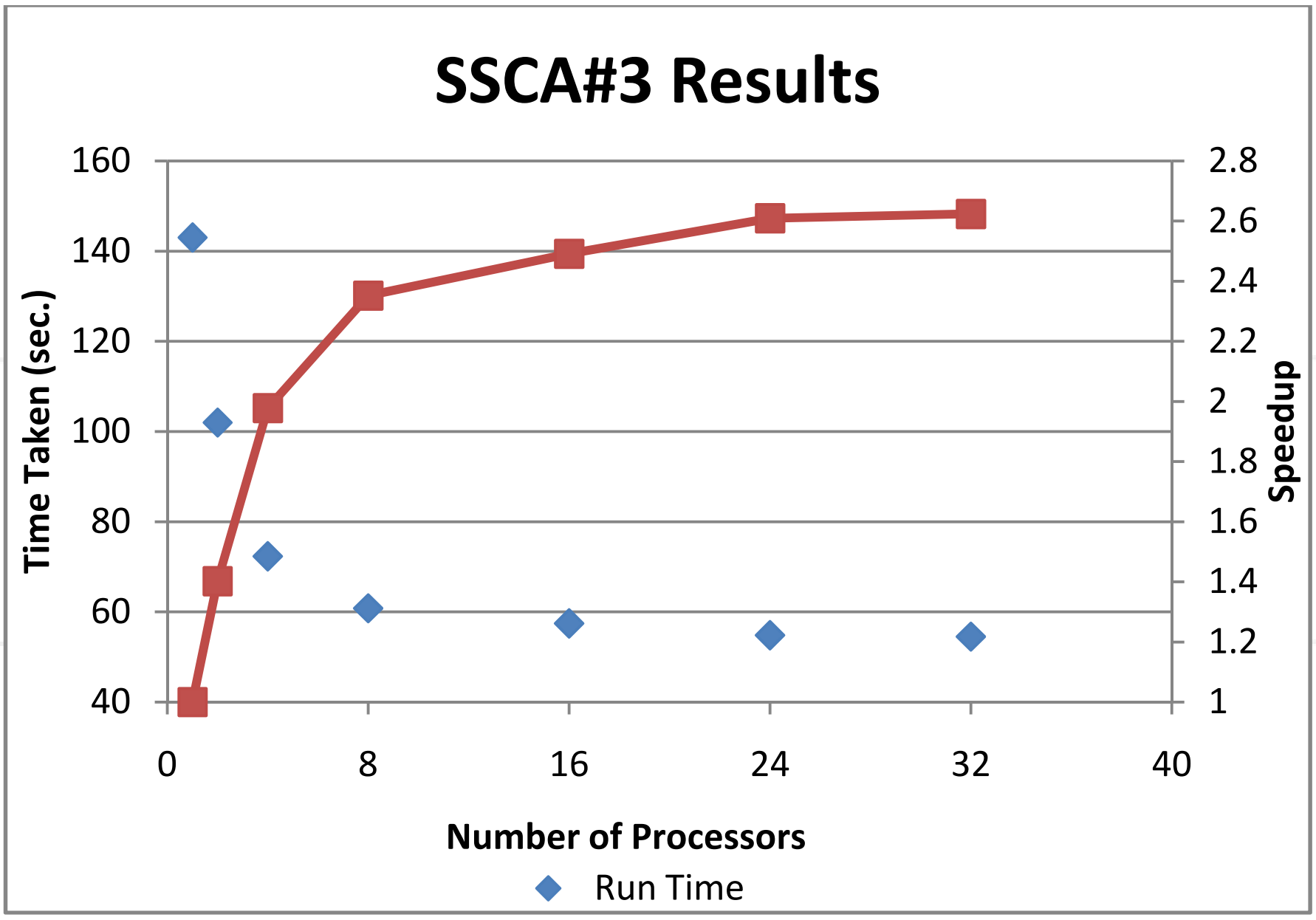

Fig. 10. SSCA#3 Results

#### 4.2.3 Results:

The pMATLAB+bcMPI implementation of the SSCA#3 benchmark was run on an AMD Opteron cluster at the Ohio Supercomputer Center with nodes containing dual 2.2 GHz Opteron processors, 4 GB RAM and an InfiniBand interconnection network. A matrix size of 1492x2296 elements was chosen, and the runs were conducted on 1, 2, 4, 8, 16, 24, and 32 processor cores. The absolute performance times and relative speedups for image formation are given in Figure 10. The graph is presented as in the previous application.
Amdahl's law states that the maximum speedup of a parallel application is inversely proportional to the percentage of time spent in sequential execution. Thus, according to Amdahl's Law, the maximum speedup possible when parallelizing 67.5% of the code is approximately 3. In the above figure, a maximum speedup of approximately 2.6 is obtained for the 32 processor run.

### 4.3 Superconducting Quantum Interference Filters (SQIF)

#### 4.3.1 Application background:

The computationally intensive signal and image processing application for this project is the modelling and simulation of Superconducting Quantum Interference Filters (SQIF) provided by researchers at SPAWAR Systems Center PACIFIC. Superconducting Quantum Interference Devices (SQUIDs) (a superconducting circuit based on Josephson junctions) and arrays of SQUIDs or Superconducting Quantum Interference Filters (SQIF) have a wide variety of applications (Palacios et al., 2006). SQUIDs are the world's most sensitive detectors of magnetic signals (sensitivity ~femto-Teslas) for the detection and characterization of signals so small as to be virtually immeasurable by any other known sensor technology. Applications such as detection of deeply buried facilities from space (military labs, WMD, etc), detection of weak signals on noise limited environments, deployment on mobile platforms, SQUID-based gravity gradiometry for navigation of submarines, biomagnetism (magnetoencephalography (MEG) and magnetocardiogram (MCG)) imaging for medical applications, detection of weapons/contraband concealed by clothing (hot spot microbolometers) and non-destructive evaluation are some of the applications based on SQUID and SQIF technologies.
Parallelization of codes that simulate SQUIDs/SQIFs are becoming extremely important for researchers. The SQIF application is intended to solve large scale SQIF problems for the study and characterization of interference patterns, flux-to-voltage transfer functions, and parameter spread robustness for SQIF loop size configurations and SQIF array fault tolerance. The SQIF application is intended to solve large scale problems relating to the field of cooperative dynamics in coupled noisy dynamical systems near a critical point. The technical background for the SQIF program can be found in (Antonio Palacios, 2006). The particular application developed was intended to run the SQIF program in an optimized fashion to either (1) reduce runtime and/or (2) increase the size of the dataset.

#### 4.3.2 Parallelization strategy:

The MATLAB profiler was used on the supplied SQIF application to determine a course of action. Application of the MATLAB profiler on the supplied *dynamics_sqif()* function using 100 SQUIDs yielded a runtime of approximately 20 minutes. A detailed analysis showed most (approximately 88%) of the time spent in the *coupled_squid()* function. Further review of the profiler results showed a linear increase in the time taken by *dynamics_sqif()* as the number of SQUIDs (*Nsquid*) was increased. Parallelization was carried out on the *dynamics_sqif()* function.

Parallelization of code consisted of adding parallel constructs to the given SQIF application. These constructs allowed the program to be run on multiple CPUs. In the course of parallelization, developers noticed the existence of task based and data based operations in the application. Task based operations were parallelized through an embarrassingly parallel (**EP**) implementation. The data based operations were parallelized through a fine grained parallel (**FP**) solution. The application was parallelized using both of the following techniques.

1. Task Based Parallelization (TP) – Parallelization over *length(xe)*
2. Data Based Parallelization (DP) – Parallelization over *Nsquid*s

A brief technical background and relative merits and demerits of each implementation are given below. It is interesting to note the difference in performance for both techniques. Results of the optimization and parallelization were obtained using pMATLAB+bcMPI on the Ohio Supercomputer Center's AMD Opteron Cluster "Glenn."

Task Parallel Approach

This particular implementation involved a task based parallel solution. The type of parallelization implemented was embarrassingly parallel. Specifically, for this application, an approach was taken such that individual processors would perform a part of the entire task, thus making the approach task parallel in nature. The embarrassingly parallel solution, in the context of this application, involved the distribution of workload between processors for the number of points for flux calculation (*length(xe)*). For this particular implementation, parallelization efforts were carried out in the *dynamics_sqif()* function. It was determined that iterations of the loop containing "*for i = 1:length(xe)*" were independent of each other. Thus, it was determined that a number of CPUs could process different portions of the 'for' loop. For example, if there were 4 processors and *length(xe)* = 100, the loop would run such that on processor 1, *i* = 1:25, on processor 2, *i* = 26:50, etc. Thus, the approach is embarrassingly parallel in nature.

A code snippet of the added pMATLAB lines required for parallelization is shown in the following figure.

```
%%%%SERIAL CODE%%%%

For i = 1:length(xe)

x = series_sqif(J(j),xe(i),M,dt,
    beta_n,Nsquid,var_size,tmax);

end
```

```
%%%%PARALLEL CODE%%%%

DVmap = map([1 Ncpus], {}, [0:Ncpus-1]);
pDV = zeros(Nsquid, length(xe), DVmap);
lngth = length(xe);
DVlocal = local(pDV);
size(DVlocal)
ind = zeros(1,Ncpus);
ind(:) = ceil(double(lngth)/Ncpus);
ind(1:rem(lngth,Ncpus)) = ceil(double(lngth)/Ncpus);
ind(Ncpus) = (ind(Ncpus)-(sum(ind)-lngth));
num_loop = ind(rank+1);
t = sum(ind(1:(rank+1)));
startind = t-ind(rank+1)+1;
endind = startind+num_loop -1;

for i = startind:endind
x = series_sqif(J(j),xe(i),M,dt,
    beta_n,Nsquid,var_size,tmax);
end
```

Fig. 11. pMatlab additions to serial code for Task Parallel Implementation

Data Parallel Approach

In this approach, a fine grained parallel solution is implemented. In the *dynamics_sqif()* function, there is a function call for *series_sqif()*. This function, *series_sqif()*, in turn calls

*my_ode()* which in turn calls *coupled_squid()*. As has been mentioned in the application background, the majority of time is spent in the *coupled_squid()* function, due to the number of times that the *coupled_squid()* function is called. The function *coupled_squid()* creates, as an output, a matrix of size *1 x Nsquid*. Looking at this, it was decided that parallelizing over *Nsquid*s would yield improvement in overall runtime. It was observed that by making suitable modifications to *coupled_squid()*, it would be possible for different processors to work on different parts of the overall data. After creating a version of *coupled_squid()* that would allow different processors to access different parts of the data, all referring function calls were modified to make use of this parallel behavior. Thus, at the topmost level, in *dynamics_sqif()*, the function *series_sqif()* could be called by different processors with different sections of the original data. A snippet of the pMATLAB code required in the *coupled_squid()* function is shown in the following figure.

```
%%%%SERIAL CODE%%%%                %%%%PARALLEL CODE%%%%

                                   xmap = map([1 Ncpus], {}, [0:Ncpus-1]);
                                   x = zeros(10001, 2*NsquidsOrig, xmap);
                                   xlocal = local(x);
x = series_sqif(J(j),xe(i),M,      xlocaltmp = series_sqif(J(j),xe(i),M,dt,beta_n, Nsquid,var_size,tmax,
    t,beta_n,Nsquid,var_size,tmax);             localpart, NsquidsOrig);
                                   xlocal = xlocaltmp(:,locPart_x);
                                   x = put_local(x, xlocal);
                                   x = agg(x);
```

Fig. 12. pMatlab additions to serial code for Data Parallel Implementation

### 4.3.3 Results

This section discusses the results of parallelizing the SQIF application by both techniques (Task parallel and Data parallel techniques). A brief discussion about the results obtained using both techniques is also presented. In the following graphs, the primary (left) vertical axis corresponds to the total time taken by the SQIF application function to complete analysis on a fixed *NSquids*. The secondary (right) axis displays the speedup obtained when running on multiple processors.

<u>Task Parallel Approach</u>

Results for running the task parallel implementation of the SQIF application were obtained on the Ohio Supercomputer Center's AMD Opteron Cluster (glenn). Near linear speedup was obtained for increasing number of processors and constant number of SQUIDs (*Nsquid*). The following graph summarizes the results obtained by running the SQIF application at OSC with a varying number of SQUIDs, and Processors.

<u>Data Parallel Approach</u>

Results for running the data parallel implementation of the SQIF application were obtained on the Ohio Supercomputer Center's AMD Opteron Cluster (glenn). A speedup was observed, and results are graphed below. The comparison is made between different numbers of SQUIDs (*Nsquid*), and different numbers of Processors. As the parallel implementation is data parallel in nature, slight modifications were made in the actual computation.

The following graph shows the application runtime and speedup for a fixed problem size (number of *Nsquid*s). A comparison is also made between the runtimes of the task parallel solution and data parallel solution.

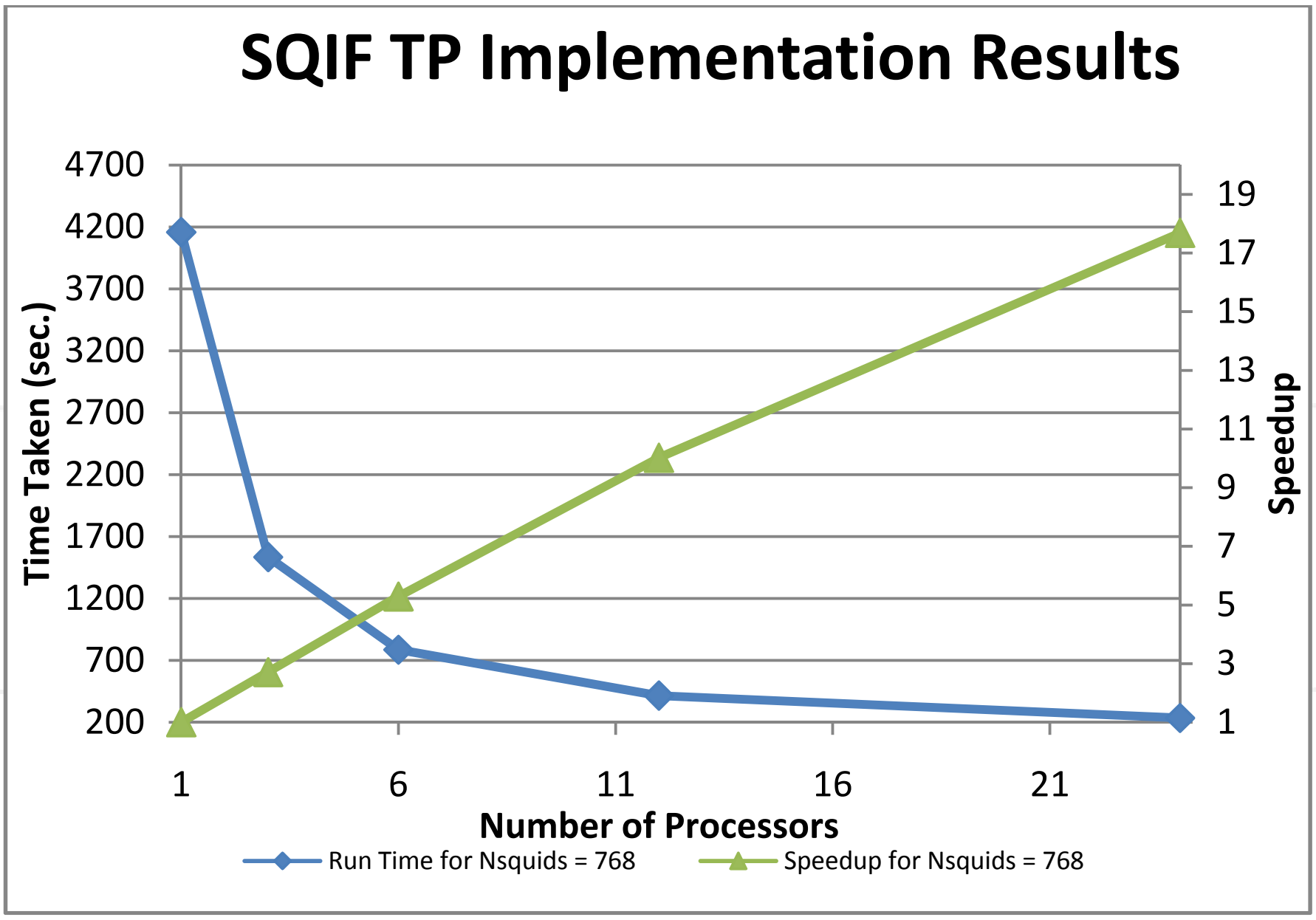

Fig. 13. Graph of TP implementation for NSquids = 768

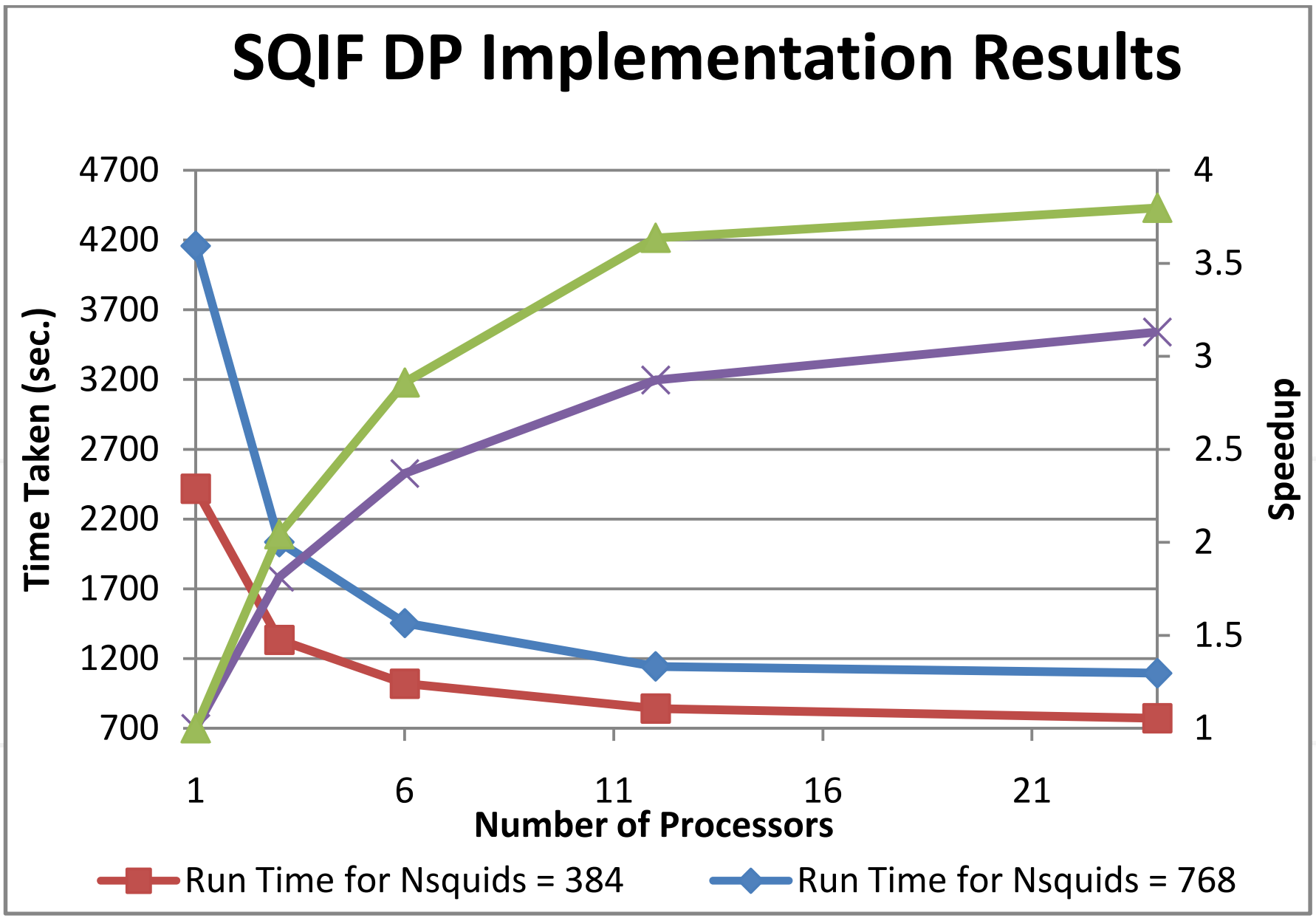

Fig. 14. Graph of DP runtimes on Glenn for Nsquids = 384 (red), 786 (blue)

From Figure 14, it is clear that there is definite speedup when using the data parallel (DP) implementation. This speedup becomes more pronounced when larger *Nsquids* are used. In these graphs, it is interesting to note that the speedup is not linear, but is much more scalable. Also, communication overhead starts to play a large part in the results when the number of processors is greater than 24 (for the problem sizes tested).
The following graph shows the performance of parallelization over *length(xe)* when compared to parallelization over *Nsquids* for a constant *length(xe)* of 100 and varying *Nsquids*. This comparison was made on the Ohio Supercomputer Center's AMD Opteron Cluster "Glenn."
From Figure 15, it is clear that parallelization over the *length(xe)* yields far better performance than parallelization over *NSquids*. This is due to the following reasons:

1. Parallelization over *length(xe)* is embarrassingly parallel. It is expected that this approach gives near linear speedup.
2. In the parallelization over *length(xe)*, the percentage of parallelized code is nearly 99%. In the parallelization over *Nsquids*, the percentage of parallelized code is nearly 90%.

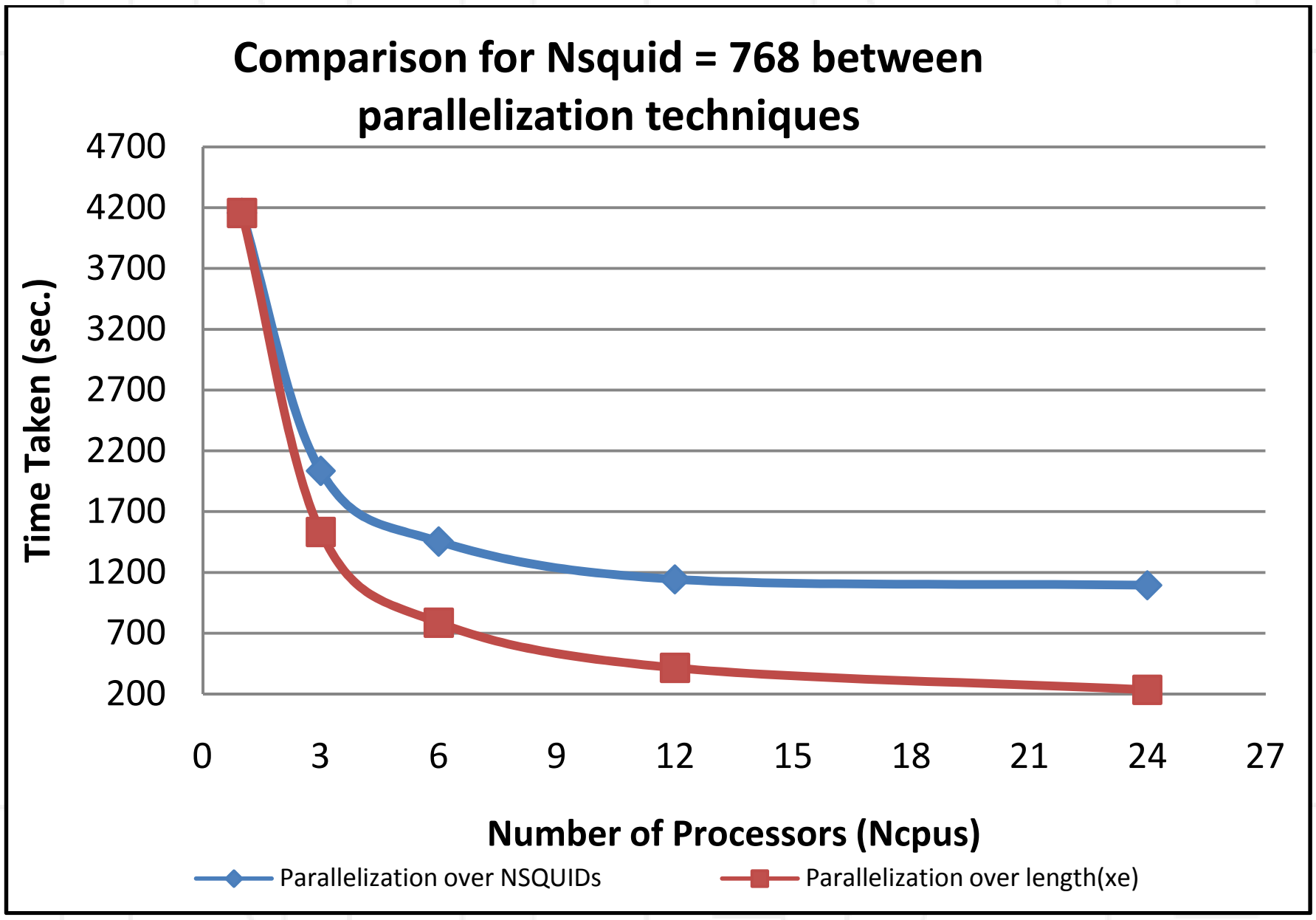

Fig. 15. Comparison between two techniques of parallelization for Nsquids = 768

Maximum speed up for both parallelization techniques:

1. By Amdahl's Law, given that parallelization is being applied to approximately 90% of the problem size, the maximum speed up one can expect is 1/(1-0.90) ≈ 10, in the case of parallelization over *Nsquids*. The maximum observed speedup is approximately 8.7 (for *Nsquids* = 14400, Number of Processors = 48).
2. By Amdahl's Law, given that parallelization is being applied to approximately 99% of the problem size, the maximum speed up one can expect is 1/(1-0.99) ≈ 100, in the case of parallelization over the *length(xe)*. The maximum observed speedup is approximately 20 (for *Nsquids* = 768, Number of Processors = 24).

## 5. Defining a general parallelization strategy:

In this section, a very general strategy for determining a parallelization strategy is presented. Please note that this strategy is very general in nature due to the numerous types of applications that can be parallelized.

1. **Determining what to parallelize:**
   Often, a user may use a tool such as the MATLAB profiler to determine this. The root cause of the slowdown needs to be determined. For example, if *function1()* is shown to be causing the slowdown, the lines of code within *function1()* that are causing the problem should be parallelized. If the user wants to parallelize the application for improved memory utilization, the source of memory usage should be determined, and this should be parallelized.
2. **Determining the type of parallelization:**
   In this step, the cause is analyzed by looking at data dependencies to determine whether an embarrassingly parallel strategy can be employed or whether a fine grained parallel strategy is required. In general, when parallelizing a large portion of code, embarrassingly parallel solutions are easier to code and deliver greater speedup when compared with fine grained parallel solutions. On the other hand, fine grained parallel solutions are useful when attempting to improve memory utilization in the application, as fine grained parallel solutions consist of data parallelization. Application of Amdahl's Law would also be beneficial, so that the developer understands what speedup to expect. For very few applications, parallelization may not lead to a significant speedup.
3. **Using one of the mentioned technologies to parallelize the application:**
   In our experience, all the mentioned technologies offer similar performance and usability for embarrassingly parallel applications. For fine-grained parallel applications, the user needs to look at the technologies more closely.
4. **Parallelize the application:**
   Recode the application with parallel constructs.
5. **Test the Application:**
   Verify that the parallelization gives correct results (often within a margin of error). As parallelization often modifies the calculations, the user needs to confirm that the parallelized code not only brings about a speedup or larger memory availability but also maintains the correct solution.

## 6. Conclusions and future work

This chapter begins with an introduction to Parallel MATLAB and its uses. From our experience, most users who require parallel MATLAB are (1) compute and/or (2) memory bound. Compute bound users often require faster time-to-solution from their MATLAB applications. Memory bound users often require the shared resources offered by using multiple processing units (more RAM, etc.). Both of these classes of users can make extensive use of parallel MATLAB technologies. Broadly, there are two techniques to parallelization (1) Task parallel (Embarrassingly parallel) or (2) Data parallel (Fine Grained parallel). Both of these techniques were described in detail, and the strategies involved with recoding an application to reflect these techniques was discussed. Three applications from the signal and image processing area were highlighted. These applications were intended to show potential users the power of parallel MATLAB, and the ease of use. Very often, for less than 5% increase in Source Lines of Code, an application can be parallelized. The applications also intended to demonstrate typical results that can be obtained by parallelizing applications using the discussed techniques. The acoustic signal processing application was parallelized using task parallel techniques, and the SSCA #3 application was parallelized using data parallel techniques. As a final application, the authors parallelized the SQIF application using both task and data parallel techniques, so demonstrate the difference between the techniques.

At the Ohio Supercomputer Center, we have had extensive experience with parallel MATLAB technologies pertaining to the Signal and Image processing area. Three parallel MATLAB technologies stand out in terms of development status: (1) bcMPI + pMATLAB (2) MATLAB DCS and Parallel Computing Toolbox, and (3) Star-P. In our experience all three technologies are equally usable, though developer preference and developer experience may play a part.

As multi-core and multi-processor systems become more common, parallel MATLAB clusters will also become more popular. MATLAB computations will be extended to Graphical Processing Units (GPUs) to harness their fast floating point arithmetic capabilities.

## 7. Acknowledgements

The authors would like to thank Gene Whipps (Army Research Laboratory, Adephi, MA, USA) for providing the ARL GRAPE application. The authors would also like to thank Dr. Fernando Escobar (SPAWAR Systems Center PACIFIC, Code 7113, San Diego, CA) for providing the serial SQIF code. The original SQIF application code was written by Dr. Patrick Loghini (SPAWAR Systems Center PACIFIC, Code 7173, San Diego, CA).

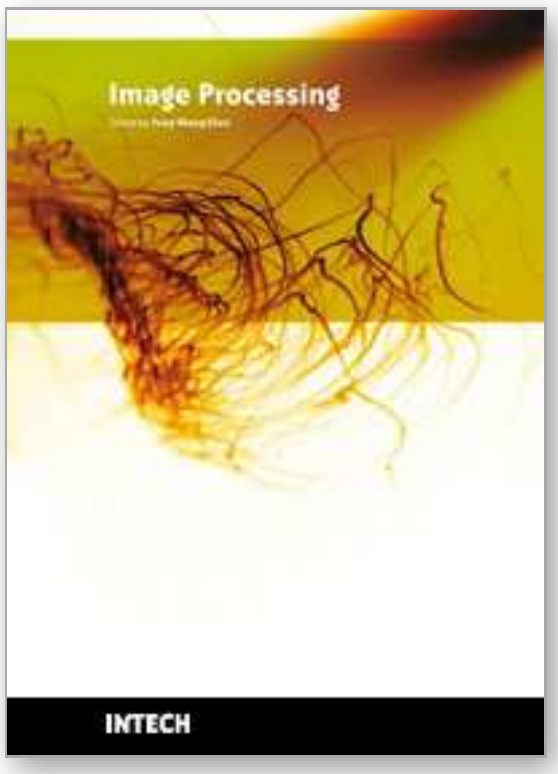

There are six sections in this book. The first section presents basic image processing techniques, such as image acquisition, storage, retrieval, transformation, filtering, and parallel computing. Then, some applications, such as road sign recognition, air quality monitoring, remote sensed image analysis, and diagnosis of industrial parts are considered. Subsequently, the application of image processing for the special eye examination and a newly three-dimensional digital camera are introduced. On the other hand, the section of medical imaging will show the applications of nuclear imaging, ultrasound imaging, and biology. The section of neural fuzzy presents the topics of image recognition, self-learning, image restoration, as well as evolutionary. The final section will show how to implement the hardware design based on the SoC or FPGA to accelerate image processing.